\documentclass[11pt,letterpaper,english,final,aps,prc,tightenlines,superscriptaddress]{revtex4}
\usepackage[T1]{fontenc}
\usepackage[latin1]{inputenc}
\usepackage{amsmath}
\usepackage{graphicx}
\usepackage{amssymb}

\makeatletter

\usepackage{geometry}

\usepackage{array}

\usepackage{longtable}

\usepackage{float}

\makeatletter

\usepackage{amsfonts}

\makeatother

\makeatother

\usepackage{babel}

\begin{document}

\title{Computer Based Analytical Simulations of the Chiral Hadronic Processes }

\author{A. Aleksejevs}

\affiliation{Division of Science, SWGC, Memorial University, Corner Brook, NL,
Canada}

\affiliation{Astronomy and Physics Department, Saint Mary's University, Halifax,
NS, Canada}

\author{M. Butler}

\affiliation{Astronomy and Physics Department, Saint Mary's University, Halifax,
NS, Canada}

\begin{abstract}
The availability of computational modeling tools for subatomic physics
(Form, FeynArts, FormCalc, and FeynCalc) has made it possible to perform
sophisticated calculations in perturbative quantum field theory. We
have adapted these packages in order to apply them to the effective
chiral field theory of hadronic interactions. A detailed description
of this Computational Hadronic Model is presented here, along with
sample calculations. 
\end{abstract}
\maketitle

\section{Introduction}

Chiral Perturbation Theory (ChPT) has been tremendously successful
in describing low-energy hadronic interactions in the non-perturbative
regime of QCD. It provides for a systematic and well-defined perturbative
expansion incorporating interaction physics driven by the symmetries
of QCD in kinematic regimes where scales are well-defined.

Calculations have been performed for many problems to Next-to-Leading-Order
(NLO) in a variety of bases, but the challenge is often in determining
the range of valid and contributing degrees of freedom to a given
problem.

With recent developments in the automatization of the NLO calculations
in perturbative field theory, it is reasonable to consider the possibility
that these methods could also be applied to ChPT and thus allow for
a broader application of the theory.

To date, there are several packages available (FeynArts \citep{FeynArts},
FormCalc \citep{FormCalc}, FeynCalc \citep{FeynCalc} and Form \citep{Form})
allowing us to produce semi-automatic calculations in particle physics.
These packages have been applied in various calculations relevant
to studies of the Standard Model \citep{BeD94,DeDH97,DeH98,GuHK99,HoS99,Ha00}.

Although FeynArts and FormCalc were originally designed for Standard
Model calculations, the flexibility of the programs allows us to extend
them to interactions appropriate for the hadronic sector. This was
a main reason to use FeynArts and FormCalc as a base languages for
the automatization of chiral hadronic calculations. Hence the main
purpose of this work is to develop an extension to FeynArts and FormCalc
which will perform automatic analytical calculations in ChPT. 

We will begin by presenting a summary of the properties of Chiral
Perturbation Theory, followed by a description of how the interactions
can be incorporated into a computational hadronic model. We will then
present some sample applications and tests of the model, along with
plans for future development.

\section{Effective Chiral Lagrangian}

Let us start with a general review of the formalism we use to describe
the dynamics of the strong interactions. Spontaneous symmetry breaking
of $SU\left(3\right)_{L}\otimes SU\left(3\right)_{R}$ due to the
pseudo-Goldstone scalar bosons into $SU\left(3\right)_{V}$ can be
described by the Lagrangian from \begin{equation}
\mathfrak{L_{\pi\pi}^{(8)}}=\frac{f_{\pi}^{2}}{8}Tr\left[D^{\mu}{\textstyle \sum\nolimits ^{\dagger}}D_{\mu}{\textstyle \sum}\right].\label{a00}\end{equation}
 Eq.(\ref{a00}) represents the first term of the effective Lagrangian
from \citep{Manohar1991}, which is the only allowed term with two
derivatives. And additional term $Tr{\textstyle \sum\nolimits ^{\dagger}}{\textstyle \sum}$
is a constant due to the equivalence theorem and is excluded from
the effective Lagrangian. Here, $f_{\pi}\approx135\, MeV$ is the
pion decay constant and ${\textstyle \sum}$ field is given by \begin{equation}
{\textstyle \sum}=e^{2iP/f_{\pi}},\end{equation}
 where $P$ is the pseudo-Goldstone boson octet: \begin{equation}
P=\left(\begin{array}{ccc}
\frac{1}{\sqrt{6}}\eta+\frac{1}{\sqrt{2}}\pi & \pi^{+} & K^{+}\\
\pi^{-} & \frac{1}{\sqrt{6}}\eta-\frac{1}{\sqrt{2}}\pi & K^{0}\\
K^{-} & \bar{K}^{0} & -\frac{2}{\sqrt{6}}\eta\end{array}\right).\label{a01}\end{equation}
 Taking into account $U\left(1\right)_{Q}$ of electromagnetism, the
covariant derivative is given by \begin{equation}
D_{\mu}=\partial_{\mu}+i\mathcal{A}_{\mu}\left[Q,...\right],\end{equation}
 where $\mathcal{A}$ enters as the electromagnetic vector field potential
and $Q$ is a charge operator. Unlike the baryon field, chiral symmetry
transformation for the pseudo-Goldstone boson field exists and is
well defined as ${\textstyle \sum}\rightarrow L{\textstyle \sum}R^{\dagger}$.
To introduce baryon field into the effective Lagrangian uniquely,
the new field $\xi$ is defined as $\xi^{2}={\textstyle \sum}$. The
chiral symmetry transformations for the field $\xi$ can be determined
in the a new basis $U$ (see \citep{Manohar1993}) as \begin{equation}
\xi\rightarrow L\xi U^{\dagger}=U\xi R^{\dagger}.\label{a1}\end{equation}
 Here, the unitary matrix $U$ is implicitly defined from Eq.(\ref{a1})
in terms of $L,\, R,\,$and $\xi$, and in this case the chiral transformation
for the baryon field is unique, and chosen to be $B\rightarrow UBU^{\dagger}.$
Choice of the basis $U$ is preferable for the effective Lagrangian
with baryons, because in this basis pions are coupled through the
derivative type coupling only. Consequently, the effective Lagrangian
for the baryons can be written in terms of the vector fields (vector
$V_{\mu}$ and axial-vector $A_{\mu}$): \begin{align}
V_{\mu} & =\frac{1}{2}\left(\xi D_{\mu}\xi^{\dagger}+\xi^{\dagger}D_{\mu}\xi\right)=\frac{1}{f_{\pi}^{2}}\left[P,\partial_{\mu}P\right]+\frac{1}{f_{\pi}^{4}}\left[P,P\left(\partial_{\mu}P\right)P\right]+...,\nonumber \\
\\A_{\mu} & =\frac{i}{2}\left(\xi D_{\mu}\xi^{\dagger}-\xi^{\dagger}D_{\mu}\xi\right)=\frac{1}{f_{\pi}}\partial_{\mu}P+\frac{1}{f_{\pi}^{3}}P\left(\partial_{\mu}P\right)P-\frac{1}{2f_{\pi}^{3}}\left\{ \partial_{\mu}P,P^{2}\right\} .\nonumber \end{align}
 The leading order baryon Lagrangian is given by \citep{Manohar1993}
\begin{equation}
\mathfrak{L}_{B\pi}^{(8)}=-iTr\,\bar{B}\mathfrak{\not D}B+m_{B}Tr\,\bar{B}B+2D\, Tr\,\bar{B}\gamma^{\mu}\gamma_{5}\left\{ A_{\mu},B\right\} +2F\, Tr\,\bar{B}\gamma^{\mu}\gamma_{5}\left[A_{\mu},B\right],\label{a2}\end{equation}
 where $B$ is $SU\left(3\right)$ octet of baryons given by \begin{equation}
B=\left(\begin{array}{ccc}
\frac{1}{\sqrt{2}}\sum^{0}+\frac{1}{\sqrt{6}}\Lambda & \sum^{+} & p\\
\sum^{-} & -\frac{1}{\sqrt{2}}\sum^{0}+\frac{1}{\sqrt{6}}\Lambda & n\\
\Xi^{-} & \Xi^{0} & -\frac{2}{\sqrt{6}}\Lambda\end{array}\right),\label{a2B}\end{equation}
 with covariant derivative defined as $\mathfrak{\not D=}\partial_{\mu}+[V_{\mu},...].$
The strong coupling constants $\left\{ F,D\right\} $ of the Lagrangian
from Eq.(\ref{a2}) have been determined in \citep{Manohar1991} to
be $F=0.40\pm0.03$ and $D=0.61\pm0.04$. At the lowest order of chiral
perturbation theory, the strong interaction of the decuplet $T_{\mu}$$\left(J^{P}=3/2^{\pm}\right)$
can be described by the following Lagrangian: \begin{equation}
\mathfrak{L}_{T\pi}^{(10)}=-i\bar{T}^{\mu}\mathfrak{\not D}T_{\mu}+m_{T}\bar{T}^{\mu}T_{\mu}+\mathcal{C}\left(\bar{B}A_{\mu}\Gamma_{5}T^{\mu}+\bar{T}^{\mu}A_{\mu}\Gamma_{5}B\right)+2\mathcal{H}\bar{T}^{\mu}\gamma^{\nu}\gamma_{5}A_{\nu}\Gamma_{5}T_{\mu},\label{a3}\end{equation}
 where $\Gamma_{5}=1$ for $T_{\mu}^{J^{P}=3/2^{+}}$ and $\Gamma_{5}=\gamma_{5}$
for $T_{\mu}^{J^{P}=3/2^{-}}$. The decuplet states $T_{\mu}$ are
given by \begin{align}
T^{333} & =\Omega^{-},\nonumber \\
\nonumber \\T^{133} & =\frac{1}{\sqrt{3}}\Xi^{\ast0},\, T^{233}=\frac{1}{\sqrt{3}}\Xi^{\ast-},\nonumber \\
\nonumber \\T^{113} & =\frac{1}{\sqrt{3}}\sum\nolimits ^{\ast+},\, T^{123}=\frac{1}{\sqrt{6}}\sum\nolimits ^{\ast0},\, T^{223}=\frac{1}{\sqrt{3}}\sum\nolimits ^{\ast-},\label{a3T}\\
\nonumber \\T^{111} & =\bigtriangleup^{++},\, T^{112}=\frac{1}{\sqrt{3}}\bigtriangleup^{+},\, T^{122}=\frac{1}{\sqrt{3}}\bigtriangleup^{0},\, T^{222}=\bigtriangleup^{-}.\nonumber \end{align}
 The strong coupling constants $\left\vert \mathcal{C}\right\vert =1.2$
and $\mathcal{H\sim}-1.8$ have been indirectly extracted from the
experimental data by \citep{Butler1993} through the loop corrections
of the strong decay of decuplet to octet of baryons in the framework
of Heavy Baryon $\chi PT$ ($HB\chi PT$) .

\section{Computational Hadronic Model}

The effective chiral theory of the strong interactions formulated
above can be applied towards a vast number of the hadronic reactions
in the nonperturbative regime. However, in this formalism, to do calculations
up to NLO by hand requires a tremendous amount of work. Because these
calculations are rather algorithmic, it is only reasonable to raise
the question whatever some degree of the automatization of the present
calculations in the ChPT is currently possible. We address this question
by developing so called Computational Hadronic Model (CHM) as an extension
of FeynArts and FormCalc packages.

The main purpose of FeynArts is to generate graphical and analytical
representations of an unevaluated amplitude for scattering or decay
processes. The results are produced by using Feynman rules as specified
in the model files and do not include any further simplifications
like index contraction or tensor decomposition and reduction. FormCalc,
although is works as a shell for Form program, can manage the explicit
analytical evaluation of the one loop integrals employing dimensional
regularization and tensor decomposition techniques. Later, the renormalization
can be performed by using either On-Shell, Minimal Subtraction ($\overline{MS}$)
or Constrained Differential Renormalization (CDR) schemes giving final
results free of ultraviolet divergences. At the final stage, an automatically
generated FORTRAN code can be used for numerical calculations of cross
section or decay widths.

Let us start with a description of the fields participating in CHM.
It is only naturally to form three classes of fields: the octet of
baryons $B$ (16 states in total with octet of pentaquarks included),
the decuplet of resonances $T^{\mu}$ (20 states in total with anti-decuplet
of pentaquark resonances included), and the octet of mesons $P$.
The external wave function for baryon (spin $J=1/2$) is given by
the usual Dirac spinor, and for the decuplet of resonances with spin
$J^{P}=3/2^{\pm}$ we use the Rarita-Schwinger (RS) field described
by

\begin{equation}
T^{\mu}\left(k,m\right)=\sum_{j=-1}^{1}\sum_{i=-1/2}^{1/2}CG_{ij,m}^{3/2}e_{j}^{\mu}\left(k\right)\cdot u\left(k,i\right)\delta_{i+j,m},\label{c1}\end{equation}
 where $CG_{ij,m}^{3/2}$ are Clebsh-Gordan coefficients defined as
\begin{equation}
CG_{ij,m}^{3/2}=\sqrt{\frac{\left(\frac{3}{2}+m\right)!\left(\frac{3}{2}-m\right)!}{3(1+j)!(1-j)!\left(\frac{1}{2}+i\right)!\left(\frac{1}{2}-i\right)!}}.\end{equation}
 The index $m$ can take values $\left\{ 3/2,1/2,-1/2,-3/2\right\} $
and describes the projections of the $J^{P}=3/2^{\pm}$ spin state.
In our model file, the field $T^{\mu}$ enters as a product of $e^{\mu}\left(k\right)\cdot u\left(k\right),$
specifying neither the polarization nor spin state. The explicit structure
of the decuplet field (see Eq.\ref{c1}) is accounted for later, when
the helicity matrix elements are computed. For the polarization vectors,
we employ the following basis in the center of mass reference frame:\begin{eqnarray}
 &  & e_{0}^{\mu}\left(k\right)=\left(\left|\overrightarrow{k}\right|,E\sin\left(\theta\right),0,E\cos\left(\theta\right)\right)/m,\nonumber \\
 &  & e_{\pm1}^{\mu}\left(k\right)=\left(0,-\cos\left(\theta\right),\mp i,-\sin\left(\theta\right)\right)/\sqrt{2}.\label{a3a}\end{eqnarray}
 Here, the angle $\theta$ determines a direction of $\overrightarrow{k}$
vector. The external wave function for the meson field $P$ is equal
to one. The spin $3/2$ propagator of the field $T^{\mu}$ is given
using convention defined in \citep{Benmerrouche}.

Since the definition of coupling matrices $\Gamma$ in the FeynArts
is based on the use of the chirality projectors $\varpi_{\pm}=\frac{1\pm\gamma_{5}}{2}$
as \begin{equation}
\Gamma=\left(\begin{array}{cc}
\mathbb{C\,}\varpi_{+}, & \mathbb{C\,}\varpi_{-}\end{array}\right)\cdot\left(\begin{array}{cc}
g_{R}^{\left(0\right)} & g_{R}^{\left(1\right)}\\
g_{L}^{\left(0\right)} & g_{L}^{\left(1\right)}\end{array}\right),\label{a4}\end{equation}
the couplings are introduced in two model files responsible for the
generic and particle levels, respectively. The first part of Eq.(\ref{a4})
describes the kinematic part of the coupling derived from the chiral
Lagrangians given by Eqs.(\ref{a00}, \ref{a2} and \ref{a3}). The
second part of the Eq.(\ref{a4}) represents the coupling strengths
defined at tree level (first column) plus counter terms (second column).
Coupling strengths $g_{R,L}^{(0)}$ in Eq.(\ref{a4}) are represented
by the Clebsh-Gordan (CG) coefficients of $SU\left(3\right)$ group
normalized by pion decay constant $f_{\pi}$ for the decays or production
channels. For the hadronic interactions, constants $g_{R,L}^{(0)}$
are given by\begin{eqnarray}
 &  & g_{R,L}^{TBP}=+\frac{\mathcal{C}}{f_{\pi}}\overline{T}^{abc}\epsilon_{ade}P_{b}^{d}B_{c}^{e},\nonumber \\
 &  & g_{R,L}^{TTP}=\pm\frac{\mathcal{H}}{f_{\pi}}\overline{T}^{abc}P_{c}^{d}T_{abd},\label{a5}\\
 &  & g_{R,L}^{BBP}=\pm\left(\frac{\left(D-F\right)}{f_{\pi}}\overline{B}_{b}^{a}P_{a}^{c}B_{c}^{b}+\frac{\left(D+F\right)}{f_{\pi}}\overline{B}_{b}^{a}B_{a}^{c}P_{c}^{b}\right),\nonumber \\
 &  & g_{R,L}^{BBPP}=+\frac{1}{f_{\pi}^{2}}.\nonumber \end{eqnarray}
 The indexes $R$ and $L$ correspond to $"+"$ and $"-"$ in the
couplings, respectively. The fields $T_{ijk}$, $B_{ij}$ and $P_{ij}$
are given by Eqs.(\ref{a01}, \ref{a2B}, \ref{a3T}), but only numerical
coefficients in front of these fields are used in Eq.(\ref{a5}).
For the $QED$ decays we have\begin{eqnarray}
 &  & g_{R,L}^{TBP\gamma}=+Q_{\pi}\frac{\mathcal{C}}{f_{\pi}}\overline{T}^{abc}\epsilon_{ade}\left[Q_{f}^{d},P_{b}^{f}\right]B_{c}^{e},\nonumber \\
 &  & g_{R,L}^{TTP\gamma}=\pm Q_{\pi}\frac{\mathcal{H}}{f_{\pi}}\overline{T}^{abc}\left[Q_{f}^{d},P_{c}^{d}\right]T_{abd},\label{a5a}\\
 &  & g_{R,L}^{BBP\gamma}=\pm\frac{Q_{\pi}}{f_{\pi}}\left(\left(D-F\right)\cdot\overline{B}_{b}^{a}\left[Q_{a}^{d},P_{d}^{c}\right]B_{c}^{b}+\left(D+F\right)\cdot\overline{B}_{b}^{a}B_{a}^{c}\left[Q_{c}^{d},P_{d}^{b}\right]\right),\nonumber \\
 &  & g_{R,L}^{BBPP\gamma}=+\frac{Q_{\pi}^{2}}{f_{\pi}^{2}},\nonumber \\
 &  & g_{R,L}^{TT\gamma}=+Q_{T},\,\, g_{R,L}^{BB\gamma}=+Q_{B},\,\, g_{R,L}^{PP\gamma}=+Q_{\pi},\,\, g_{R,L}^{PP\gamma\gamma}=+Q_{\pi}^{2}.\nonumber \end{eqnarray}
 Here $Q_{T,B,\pi}$ is a charge of $\left\{ T_{\mu},B,P\right\} $
fields given in the units of electron's charge. $SU\left(3\right)$
electromagnetic charge matrix $Q$ is defined as \begin{equation}
Q=\left(\begin{array}{ccc}
2/3 & 0 & 0\\
0 & -1/3 & 0\\
0 & 0 & -1/3\end{array}\right).\end{equation}
 The counter terms of equation Eq.(\ref{a4}) are given through the
wave function renormalization of the baryon field. Here we adopt notation
of \citep{Denner} \begin{equation}
g_{R,L}^{\left(1\right)}\left(A\rightarrow B\right)=\frac{1}{2}\left(\delta f_{R,L}^{A}+\delta f_{R,L}^{B}\right)g_{R,L}^{\left(0\right)}\left(A\rightarrow B\right),\end{equation}
 where indices $\left\{ A,B\right\} $ correspond to the external
baryons of the process $A\rightarrow B$. The wave function renormalization
constants $\left\{ \delta f_{L,R}^{A},\delta f_{L,R}^{B}\right\} $
are computed through the truncated self energy graphs $\sum$, and
given as an example for the baryons of the type $A$\begin{equation}
\delta f_{R,L}^{A}=-Re\left(\sum\nolimits _{A,A}^{R,L}\left(m_{A}^{2}\right)\right)-m_{A}^{2}\cdot Re\left(\frac{\partial}{\partial p^{2}}\left[\sum\nolimits _{A,A}^{L}\left(p^{2}\right)+\sum\nolimits _{A,A}^{R}\left(p^{2}\right)+2\sum\nolimits _{A,A}^{S}\left(p^{2}\right)\right]\right)_{p^{2}=m_{A}^{2}},\end{equation}
 where $L,R$ and $S$ are left-handed, right-handed and scalar parts
of the truncated self energy, respectively. Another counter term arising
from the radiative decays of the decuplet into the octet state $\left\{ T\rightarrow B+\gamma\right\} $
is described by the Lagrangian \begin{equation}
\mathfrak{L}^{TB\gamma}=i\Theta\frac{e}{\Lambda_{\chi}}\overline{B}\gamma^{\mu}\gamma_{5}QT^{\nu}F_{\mu\nu}.\end{equation}
 The unknown coupling constant $\Theta$ has been determined in \citep{Butler1993}
from the measured branching ratios of $\Delta\rightarrow N\gamma$
and $\Xi^{*0}\rightarrow\Xi^{0}\gamma$. The Clebsh-Gordan (CG) coefficients
for this counter term are simply given by \begin{equation}
g_{R,L}^{\left(1\right)}\left(T\rightarrow B\gamma\right)=\pm\Theta\frac{e}{\Lambda_{\chi}}\overline{T}^{abc}\epsilon_{ade}Q_{b}^{d}B_{c}^{e},\end{equation}
 and have dimension of $1/M$.

\begin{figure}
\begin{centering}
\includegraphics[scale=0.5]{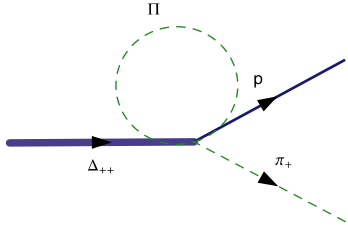} 
\par\end{centering}

\caption{Next-to-the-leading hadronic decay with the adjacency of order five
for the decay $\Delta_{++}\rightarrow p\pi_{+}.$\label{fig1}}

\end{figure}

Due to the presence of the $3\pi$ interaction in the chiral theory,
a coupling of the adjacency of order five (see, for example, Fig.\ref{fig1})
is introduced in the hadronic decays of the baryons.%

Since, the couplings of the order five adjacency (see Fig.(\ref{fig1}))
described by the same CG coefficients as for the tree level couplings
of order three adjacency, it is straightforward to program the same
coupling constants for every pion field appearing in the loop. This
can be easily accomplished if the field $\Pi$ is introduced in the
loop. This field $\Pi$ appears only in the loops with the $3\pi$
coupling, with a propagator described as \begin{equation}
\Delta_{\Pi}(q)=\sum_{i=1}^{8}\frac{1}{q^{2}-m_{i}^{2}},\end{equation}
 where $m_{i}$ is the mass of the meson in the $SU\left(3\right)$
octet (see Eq.(\ref{a01})). The kinematic part of the coupling for
this type of processes is exactly equal to the couplings of the hadronic
interactions described by chiral Lagrangians (Eqs.(\ref{a00}, \ref{a2}
and \ref{a3})). The same is true for the CG coefficients, but with
$f_{\pi}$ constant replaced by $f_{\pi}^{3}$ in the first line of
Eq.(\ref{a5}).

\section{Applications and Tests}

The extension towards the hadronic sector in FeynArts and FormCalc
which we propose in this work can be applied to the virtually any
hadronic process. We choose two processes relevant to the problems
of the hadronic physics. These are pentaquark photo-production and
Compton scattering in the studies of the nucleon polarizabilities.
Since these problems were addressed earlier by \citep{Hosaka05} and
\citep{Meissner}, this will serve as a test of the rather involved
calculations using our Computational Hadronic Model (CHM).

\subsection{Pentaquark Photo-Production}

The question of existence of exotic baryons is of crucial importance
to the physics of the strong interactions. The bound states of five
quarks (pentaquark) can be described by the at least two models. Nucleon-Kaon
molecule model, proposed by \citep{Diakonov97}, assumes that pentaquark
is formed by a nucleon and a kaon bounded via the nuclear strong interaction.
An alternative model proposed by \citep{Wilczek} treat a pentaquark
as a system of two diquarks and one anti-quark $(q^{4}\overline{q})$,
bounded via the colour strong interaction. Since the range of the
nuclear strong interaction is of order of $\sim1\, fm$, compared
to the color strong interaction range, of $\sim0.1\, fm$, the decay
widths of the pentaquark in these two models are substantially different
($\sim100\, MeV$ vs $\sim1\, MeV$). Obviously, the detection and
subsequent measurements of the pentaquark decay width will most probably
rule out one of these models. In the experimental searches of the
exotic baryons, attention goes towards lightest pentaquark state $\Theta_{5}^{+}=\{uuds\bar{s}\}$,
which has a mass around $1540\, MeV$. This state is attractive not
only because it has the highest production probability, but also due
to the fact that symmetries of the chiral Lagrangian forbid the electromagnetic
decay of this state to the proton and photon at the tree level and
hence makes $\Theta_{5}^{+}$ state more stable with decay width probably
around $1\, MeV$. Unfortunately, despite of tremendous experimental
efforts, production of this state proved to be extremely problematic.
The experimental difficulties emerge from the weak signal, low statistics,
and high background in all production channels. 

From the point of view of relativistic ChPT it is possible to introduce
pentaquark baryons in to the computational hadronic model due to the
same symmetry of the chiral Lagrangian and similarities of the $10$
and $\overline{10}$ irreducible representations of the $SU(3)$ group.
If one assumes that octet of the pentaquarks is represented by the
spin $1/2$ states, and the anti-decuplet of resonances by the spin
$J^{P}=3/2^{\pm}$, it easy to construct a Lagrangian for both. Using
notation of \citep{Beane}, we can describe the octet of exotic baryons
by \begin{equation}
\mathcal{O}_{5}=\left(\begin{array}{ccc}
\frac{1}{\sqrt{2}}\sum_{5}^{0}+\frac{1}{\sqrt{6}}\Lambda_{5} & \sum_{5}^{+} & p_{5}\\
\sum_{5}^{-} & -\frac{1}{\sqrt{2}}\sum_{5}^{0}+\frac{1}{\sqrt{6}}\Lambda_{5} & n_{5}\\
\Xi_{5}^{-} & \Xi_{5}^{0} & -\frac{2}{\sqrt{6}}\Lambda_{5}\end{array}\right),\end{equation}
 and the anti-decuplet of pentaquark resonances by\begin{align}
T_{5}^{333} & =\Theta_{5}^{+},\nonumber \\
\nonumber \\T_{5}^{133} & =\frac{1}{\sqrt{3}}N_{5}^{0},\, T_{5}^{233}=\frac{1}{\sqrt{3}}N_{5}^{+},\nonumber \\
\nonumber \\T_{5}^{113} & =\frac{1}{\sqrt{3}}\sum\nolimits _{5}^{-},\, T_{5}^{123}=\frac{1}{\sqrt{6}}\sum\nolimits _{5}^{0},\, T_{5}^{223}=\frac{1}{\sqrt{3}}\sum\nolimits _{5}^{+},\\
\nonumber \\T_{5}^{111} & =\Xi_{5}^{--},\, T_{5}^{112}=\frac{1}{\sqrt{3}}\Xi_{5}^{-},\, T_{5}^{122}=\frac{1}{\sqrt{3}}\Xi_{5}^{0},\, T_{5}^{222}=\Xi_{5}^{+}.\nonumber \end{align}
 The interaction Lagrangian for all the possible permutations of the
fields $B,T_{\mu},\mathcal{O}_{5},T_{5,\mu}$ and $A_{\mu}$ can be
written in the following form: \begin{eqnarray}
\mathfrak{L}^{(\overline{10}\oplus8)}= & 2\mathcal{D}_{\mathcal{O}}\, Tr\,\bar{\mathcal{O}}_{5}\gamma^{\mu}\gamma_{5}\left\{ A_{\mu},\mathcal{O}_{5}\right\} +2\mathcal{F}_{\mathcal{O}}\, Tr\,\bar{\mathcal{O}_{5}}\gamma^{\mu}\gamma_{5}\left[A_{\mu},\mathcal{O}_{5}\right]+2\mathcal{H}_{P}\bar{T}_{5}^{\mu}\gamma^{\nu}\gamma_{5}A_{\nu}\Gamma_{5}T_{5,\mu}+\nonumber \\
 & \mathcal{C}_{P\mathcal{O}}\left(\bar{\mathcal{O}_{5}}A_{\mu}\Gamma_{5}T_{5}^{\mu}+\bar{T}_{5}^{\mu}A_{\mu}\Gamma_{5}\mathcal{O}_{5}\right)+\mathcal{C}_{PB}\left(\bar{B}A_{\mu}\Gamma_{5}T_{5}^{\mu}+\bar{T}_{5}^{\mu}A_{\mu}\Gamma_{5}B\right)+\label{eq:penta}\\
 & 2\mathcal{D}_{\mathcal{O}B}\,\left(Tr\,\bar{\mathcal{O}}_{5}\gamma^{\mu}\gamma_{5}\left\{ A_{\mu},B\right\} +h.c.\right)+2\mathcal{F}_{\mathcal{O}B}\,\left(Tr\,\bar{B}\gamma^{\mu}\gamma_{5}\left[A_{\mu},\mathcal{O}_{5}\right]+h.c.\right).\nonumber \end{eqnarray}
 Relevant to the physics of the pentaquark production, coupling constant
$\mathcal{C}_{PB}$ in Eq.(\ref{eq:penta}) is given by $\mathcal{C}_{PB}=\frac{g_{TBP}\, f_{P}}{M_{P}}$.
The constant $g_{TBP}$ can be extracted from the information on hadronic
decay width of the specific pentaquark state. The rest of the coupling
constants can either be extracted using the Adler-Weisberger sum rules
(see \citep{Beane}) or require additional experimental input. The
Clebsh-Gordan coefficients for the pentaquark states were computed
in the same way as in Eq.(\ref{a5}) and Eq.(\ref{a5a}) but with
coupling constants taken from Lagrangian in Eq.(\ref{eq:penta}).

\begin{figure}
\begin{centering}
\includegraphics[scale=0.8]{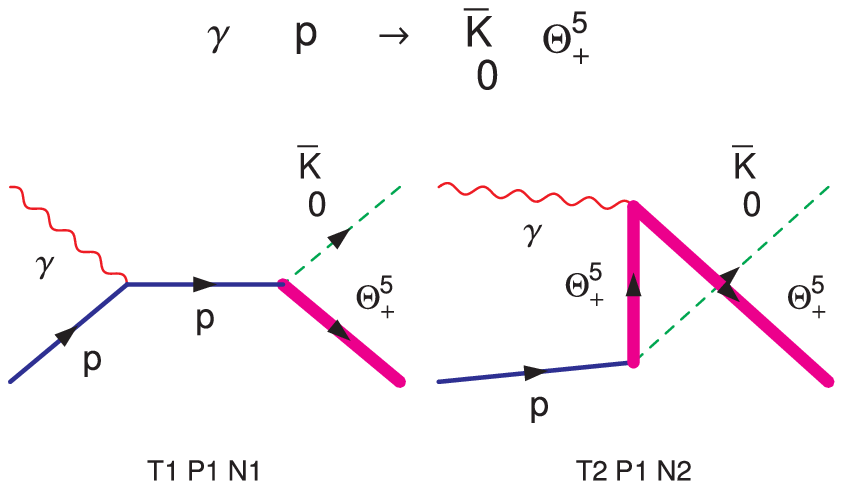} ~~~~\includegraphics[scale=0.8]{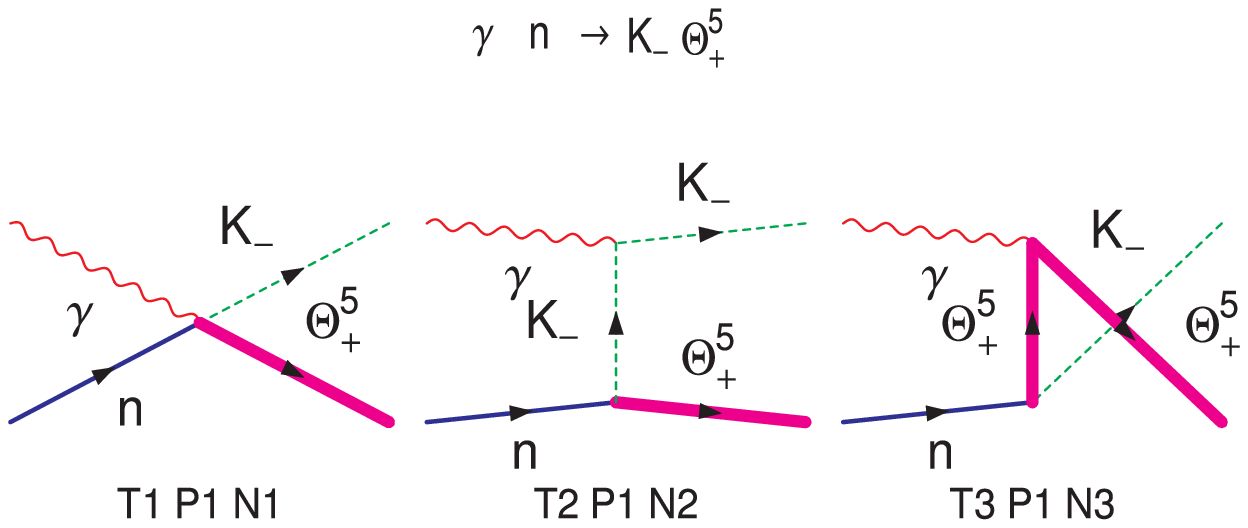} 
\par\end{centering}

\caption{$\Theta_{5}^{+}$photo-production predictions from the FeynArt's Computational
Hadronic Model.\label{fig:theta_proton}}

\end{figure}

\begin{figure}
\begin{centering}
\includegraphics[scale=0.45]{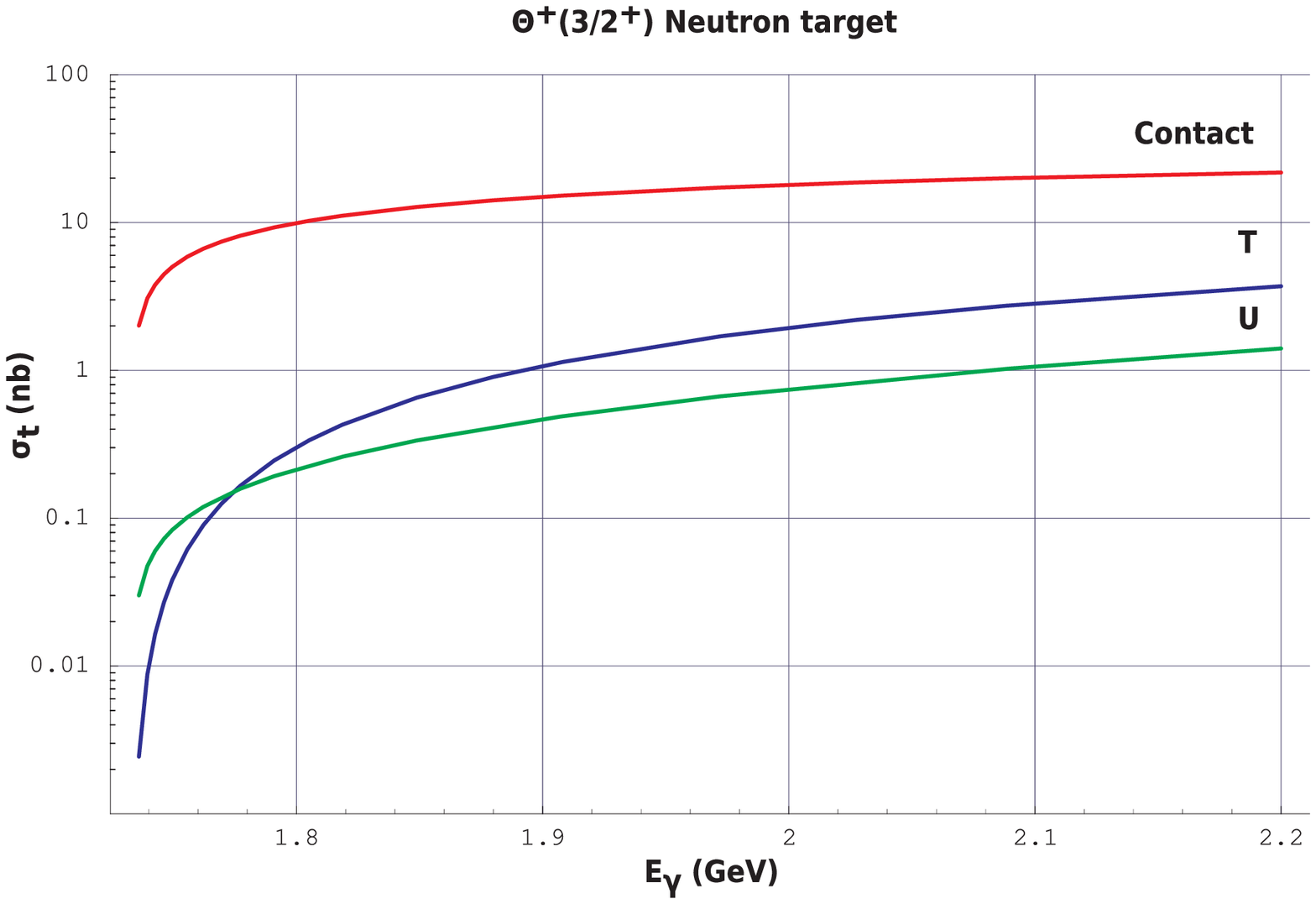} ~~~~~\includegraphics[scale=0.45]{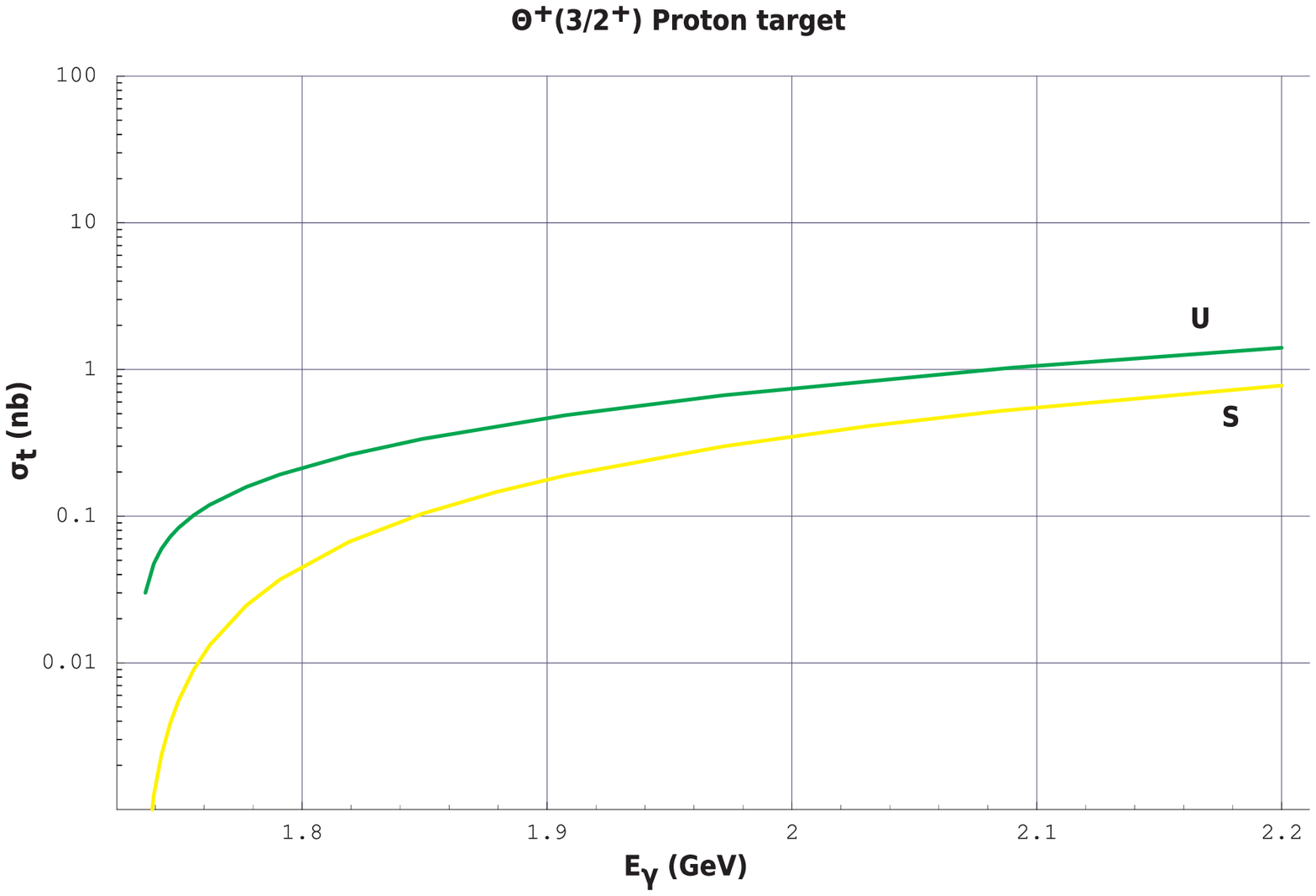} 
\par\end{centering}

\begin{centering}
\includegraphics[scale=0.45]{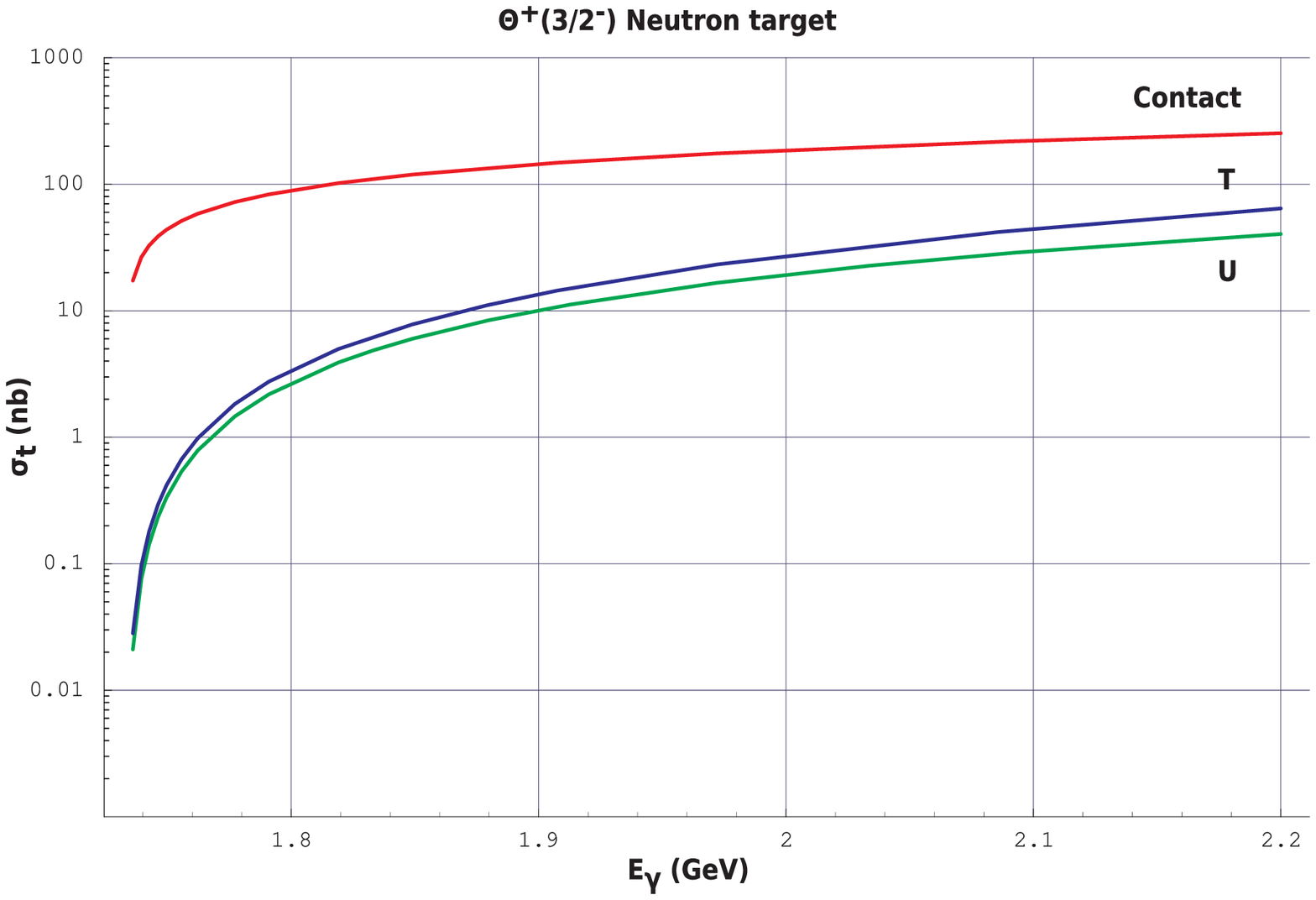}~~~~~\includegraphics[scale=0.45]{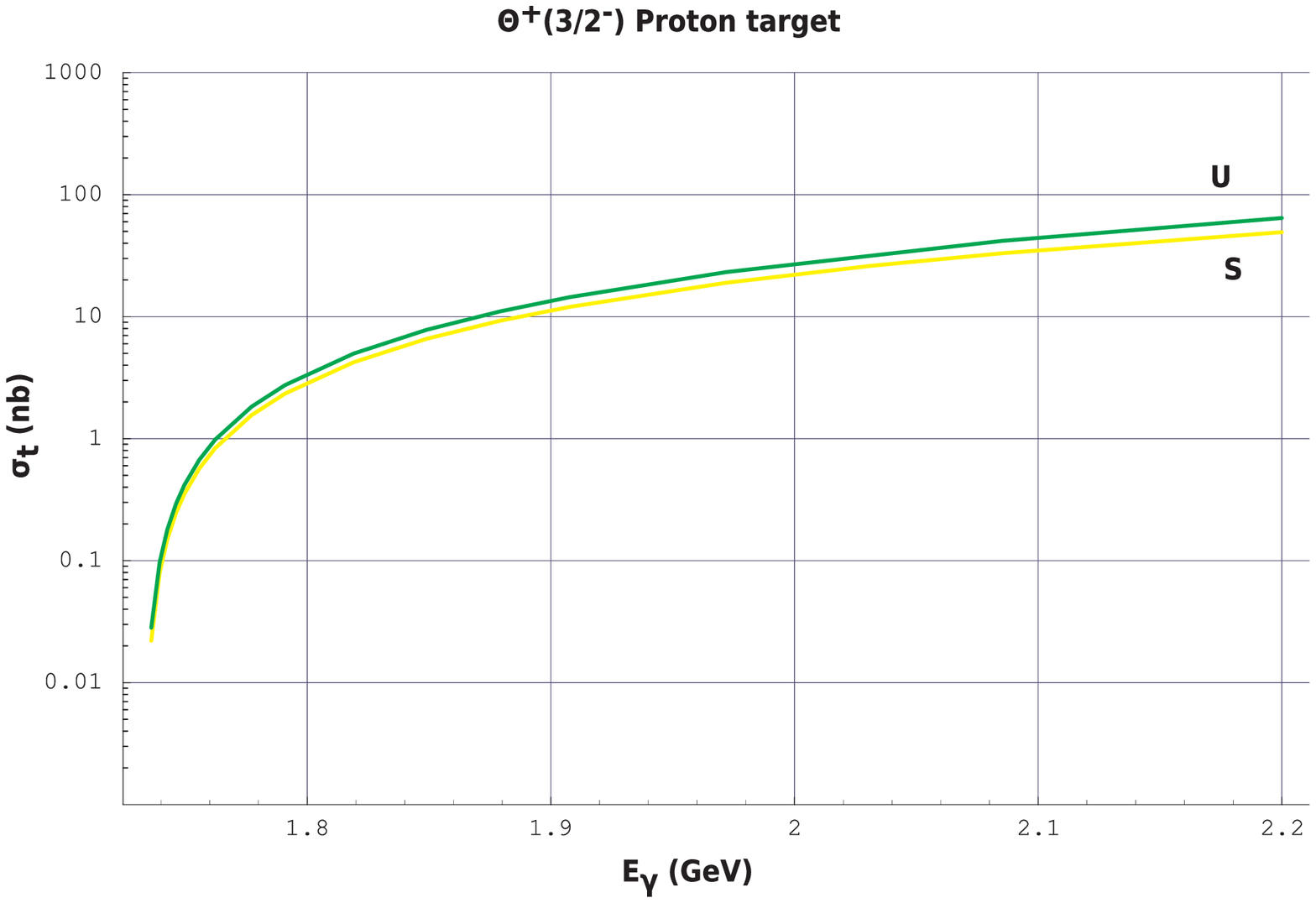} 
\par\end{centering}

\caption{The total cross section for the $\Theta_{5}^{+}$ photo-production
in the contact, s-, t- and u-channels as a function of energy of the
photon $E_{\gamma}$ in the laboratory reference frame. \label{fig:penta_photo}}

\end{figure}

It is natural to raise a question: if the $\Theta_{5}^{+}$ does exist,
how can chiral effective theory direct experimental searches for this
state? One of the possibilities would be to look at the kinematic
dependencies (see \citep{Hosaka05}) using our Computational Hadronic
Model. There are three possible channels for the production $\Theta_{5}^{+}$:
photo-, hadro- and lepto-production. The most recent experimental
efforts were directed to the photo-production of the $\Theta_{5}^{+}$.
Photo-production is described by the two reactions, $\gamma p\rightarrow\Theta_{5}^{+}\overline{K}_{0}$
on the proton target and $\gamma n\rightarrow\Theta_{5}^{+}K_{-}$
on the neutron target. The decision to look at these two channels
explicitly is motivated by controversy over experimental results (see
\citep{CLAS05}). Our goal here is to construct the production cross
sections for these two reactions and to compare our results to \citep{Hosaka05}.
In the Born approximation diagrams, responsible for the production
of $\Theta_{5}^{+}$ are given by the contact, $s$, $t$ and $u$
channels and shown on the Fig.(\ref{fig:theta_proton}).%
 Obviously there is a distinctive asymmetry in the production channels
for the two targets. This asymmetry is reflected by the presence of
the contact term in photoproduction on the neutron target. This result
was pointed out originally by \citep{Hosaka05} and confirmed in our
CHM. According to the Lagrangian given by Eq.(\ref{eq:penta}), we
can construct two parity states of $\Theta_{5}^{+}$ with $J^{P}=3/2^{+}$or
$J^{P}=3/2^{-}$ with values for the coupling constant $g_{KN\Theta}=0.53$
for $\Theta_{5}^{+}(3/2^{+})$ and $g_{KN\Theta}=4.22$ for $\Theta_{5}^{+}(3/2^{-})$
\citep{Hosaka05} if we assume that this state has a narrow decay
width around $1\, MeV.$ The production cross sections for the contact,
$s-,$ $t-$ and $u$- channels are shown on Fig.(\ref{fig:penta_photo})
as functions of the photon's energy $E_{\gamma}$ in the laboratory
reference frame.%
 The largest contribution comes from the contact term which is only
present for the neutron target. Comparing our results to \citep{Hosaka05}
we can see what s- and u-channel cross sections are somewhat larger
in our case. This can be explained by the fact that, unlike in CHM
model, the couplings in \citep{Hosaka05} are adjusted by the four-dimensional
formfactors, and, since in these channels baryons are further off-shell,
the contribution of these channels is strongly suppressed. As for
the contact and t-channel contribution, our results are close to \citep{Hosaka05}.
The evident asymmetry between neutron and proton targets favors deuterium
target (with proton as a spectator) for the photo-production of the
$\Theta_{5}^{+}(3/2^{\pm})$ with production cross section around
$\sim200\, nb$ and $\sim30\, nb$, respectively.

\subsection{Compton Scattering}

Although the CHM predictions of the pentaquark photoproduction confirmed
the earlier theoretical findings, our calculations are done using
the Born approximation only, and hence it is reasonable to raise the
question about the validity of our CHM extension of FeynArts and FormCalc
for the Next-to-the-Leading-Order (NLO) calculations. 

\begin{figure}
\begin{centering}
\includegraphics[scale=0.6]{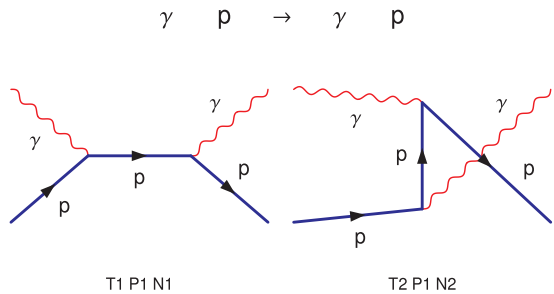} 
\par\end{centering}

\caption{FeynArts output for the Compton scattering at the tree level in CHM.\label{fig:compton-tree}}

\end{figure}

\begin{figure}
\begin{centering}
\includegraphics[scale=0.5]{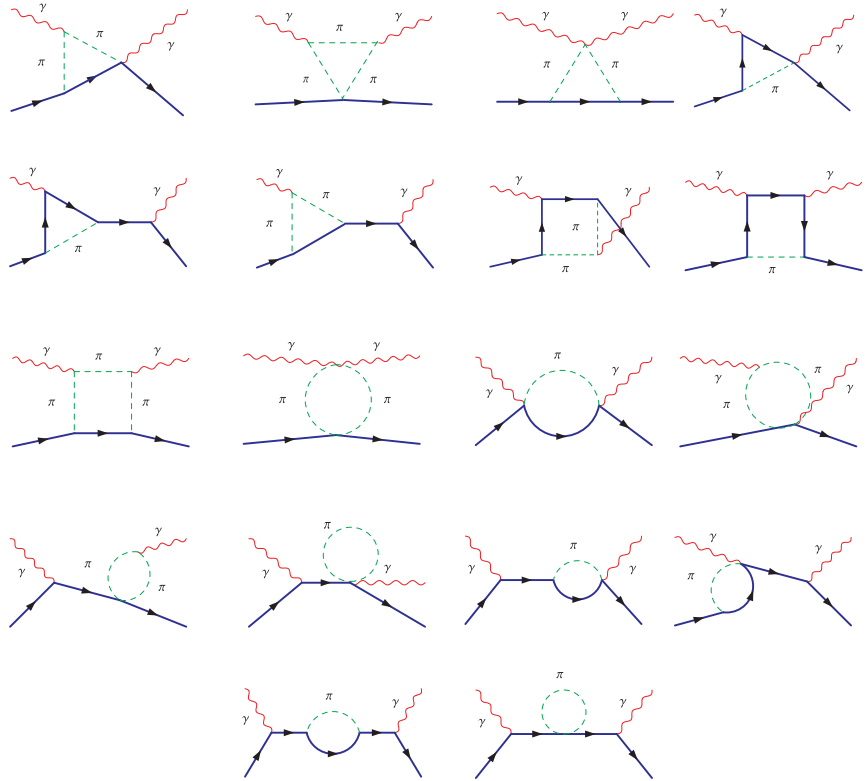} 
\par\end{centering}

\caption{Representative diagrams for the Next-to-the-Leading order Compton
scattering in CHM.\label{fig:loop-SU2}}

\end{figure}

To test our NLO calculations (not including resonances) we decided
to compute amplitudes for the Compton scattering off the proton and
neutron. If we consider a nucleon with structure, the Compton scattering
amplitude in the rest frame has the following form \citep{Meissner}:\begin{equation}
M(\gamma N\rightarrow\gamma N)=-\frac{e^{2}Z^{2}}{4\pi m}\overrightarrow{\epsilon}'\cdot\overrightarrow{\epsilon}+\alpha\omega'\omega\overrightarrow{\epsilon}'\cdot\overrightarrow{\epsilon}+\beta(\overrightarrow{\epsilon}'\times\overrightarrow{k}')(\overrightarrow{\epsilon}\times\overrightarrow{k})+\mathcal{O}(\omega^{4}).\label{a7}\end{equation}
 Here, $(\overrightarrow{\epsilon},\,\omega,\,\overrightarrow{k})$
is the polarization vector, frequency and momenta of the incoming
photon, respectively. Primed quantities denote the outgoing photon.
Two structure constants $\alpha$ and $\beta$ are the electric and
magnetic polarizabilities of the nucleon. For the point like nucleon,
we have (see Fig.(\ref{fig:compton-tree})):

\begin{equation}
M_{0}=-\frac{e^{2}Z^{2}}{4\pi m}\overrightarrow{\epsilon}'\cdot\overrightarrow{\epsilon}.\label{a6}\end{equation}
 This amplitude, in order to satisfy the gauge invariance of the electromagnetic
current $(\partial^{\mu}J_{\mu}=0),$ is not renormalized, and expressed
through the set of physical observables such as charge and mass. Moreover,
in the Thomson limit when $\overrightarrow{k}'\rightarrow\overrightarrow{k}$,
the Compton amplitude $M$ in Eq.(\ref{a7}) should take the form
of $M_{0}$ in Eq.(\ref{a6}) according to the well known low-energy
theorem. Hence, when soft photons are considered, the nucleon behaves
as a point-like particle and structure constants of Eq.(\ref{a7})
could be calculated only through the loop diagrams. Although generally
the loop contribution is ultraviolet divergent and requires renormalization,
this is not a case for the Compton scattering. If $M(\overrightarrow{k}'\rightarrow\overrightarrow{k})\rightarrow M_{0}$,
$M_{0}$ is not renormalized and the limit $\overrightarrow{k}'\rightarrow\overrightarrow{k}$
does not remove ultraviolet divergences, then the amplitude $M$ should
be finite. If we take only the $SU(2)$ triplet of the mesons, we
will have to evaluate 22 diagrams for the neutron and 52 diagrams
for the proton (see Fig.(\ref{fig:loop-SU2})).%
 It is necessary to mention that the diagrams on Fig.(\ref{fig:loop-SU2})
represent only generic types of the topologies allowed in Compton
scattering. When one employs CHM, the full set of graphs will have
to include the crossed diagrams and wave function renormalization
graphs absorbed into counterterms. Also, in the CHM, charged mesons
are treated as non-selfconjugate fields which effectively induce double
counting, so part of the final amplitude with the charged mesons in
the loops has to be divided by the factor of two.

Individually, the diagrams on Fig.(\ref{fig:loop-SU2}) give the divergent
contributions but their sum should be a finite result. That was a
case when Compton amplitude was computed in the CHM. We observed analytically
that the final Compton amplitude is, in fact, finite (i.e. free of
ultraviolet divergences). Moreover, we have successfully tested finiteness
of the Compton amplitude for the $SU(3)$ octet of mesons. Here we
have calculated 44 graphs for the neutron and 104 (not including wave
function renormalization graphs) for the proton and observed no ultraviolet
divergences in the final amplitude. In the FormCalc, the final results
(amplitude or Compton tensor) were presented analytically using the
Passarino-Veltman basis. Due to the cumbersome nature of this type
of calculations we leave analytical details out of this article. Although
we can proceed and express polarizabilities of the nucleon numerically,
we leave that for the next publication. We have satisfied purpose
of this article already when ultraviolet finite results for the Compton
amplitude were obtained up to NLO using our computational hadronic
model.

\section{Conclusion and Outlook}

In the present work we have developed an extension (the Computational
Hadronic Model) of FeynArts and FormCalc to include the hadronic sector
using Chiral Perturbation Theory. We have included octet of mesons,
baryons (and pentaquark baryons) plus the decuplet of resonances (and
the anti-decuplet of pentaquark resonances). The CHMs provides a robust
prediction of the asymmetry in pentaquark $\Theta_{5}^{+}$ photoproduction
between neutron and proton targets. It can be concluded that the asymmetry
is caused by the presence of the contact term in the $\gamma n\rightarrow K^{-}\Theta_{5}^{+}$channel.

The ability of the CHM to handle Next-to-the-Leading-Order calculations
was tested through the fact that the Compton scattering amplitude
should be ultraviolet finite and should not be renormalized. As was
expected, even when the entire octet of mesons was included, the final
amplitude did not exhibit any ultraviolet divergent behavior. The
NLO results of the CHM can be expressed analytically using FormCalc,
with an amplitude expressed in the Passarino-Veltman basis as well
as numerically with application of the package LoopTools.

Future applications of the CHM can be envisioned for calculations
of many processes, including the full kinematic dependencies for the
production/decay channels of hadronic physics. Source files of the
CHM are available by request from the authors.

\section{Acknowledgment}

This work has been supported by both an NSERC Discovery Grant, and
A. Aleksejevs is grateful for support from Saint Mary's University
through its NSERC Research Capacity Development Grant. We are grateful
to S. Barkanova of Acadia University for the useful discussions. The
authors would also like to acknowledge the role of the T. Hahn who
made packages such as FeynArts, FormCalc and LoopTools available to
physics community.

\end{document}